\documentclass[conference]{IEEEtran}
\IEEEoverridecommandlockouts

\usepackage{cite}
\usepackage{amsmath,amssymb,amsfonts}
\usepackage{algorithm,algpseudocode}
\usepackage{graphicx}
\usepackage{textcomp}
\usepackage{xcolor}
\usepackage[hidelinks]{hyperref}
\usepackage{xurl}
\usepackage{float}
\usepackage{comment}
\newcommand{\sysname}{{\scshape LoRAServe}~}
\newcommand{\sysnamex}{{\scshape LoRAServe}}
\newcommand{\orgname}{Company X~}
\newcommand{\orgnamex}{Company X}
\usepackage{tikz}
\newcommand*\circled[1]{\tikz[baseline=(char.base)]{
            \node[shape=circle,draw,inner sep=2pt] (char) {#1};}}

\newcommand{\xa}[1]{{\textcolor{orange}{[{\bfseries Chloe:} #1]}}}

\newcommand{\shashwat}[1]{{{\color{cyan} [Shashwat: #1]}}}
\newcommand{\anjaly}[1]{{{\color{teal} [Anjaly: #1]}}}

\newcommand{\ankur}[1]{{{\color{purple} [Ankur: #1]}}}

\def\BibTeX{{\rm B\kern-.05em{\sc i\kern-.025em b}\kern-.08em
    T\kern-.1667em\lower.7ex\hbox{E}\kern-.125emX}}
\begin{document}

\pdfpagewidth=8.5in
\pdfpageheight=11in

\pagenumbering{arabic}

\title{Serving Heterogeneous LoRA Adapters in Distributed LLM Inference Systems}
\author{
\IEEEauthorblockN{
Shashwat Jaiswal$^1$, 
Shrikara Arun$^2$, 
Anjaly Parayil$^2$, 
Ankur Mallick$^2$, 
Spyros Mastorakis$^2$,
Alind Khare$^2$, \\
Chloi Alverti$^3$,
Renee St Amant$^2$,
Chetan Bansal$^2$,
Victor Rühle$^2$,
Josep Torrellas$^1$
}
\IEEEauthorblockA{\{sj74, torrella\}@illinois.edu, \{t-sarun, aparayil, ankurmallick, smastorakis, \\alindkhare, reneestamant, chetanb, virueh\}@microsoft.com, xalverti@cslab.ece.ntua.gr} 
$^1$University of Illinois Urbana-Champaign, $^2$Microsoft, $^3$National Technical University of Athens
}

\maketitle
\thispagestyle{plain}
\pagestyle{plain}


\begin{abstract}
Low-Rank Adaptation (LoRA) has become the de facto method for parameter-efficient fine-tuning of large language models (LLMs), enabling rapid adaptation to diverse domains. 
In production, LoRA-based models are served at scale, creating multi-tenant environments with hundreds of adapters sharing a base model. 
However, state-of-the-art serving systems co-batch heterogeneous adapters without accounting for rank (size) variability, leading to severe performance skew, which ultimately requires adding more GPUs to satisfy service-level objectives (SLOs). Existing optimizations, focused on loading, caching, and kernel execution, ignore this heterogeneity, leaving GPU resources underutilized.
We present \sysnamex, a workload-aware dynamic adapter placement and routing framework designed to tame rank diversity in LoRA serving. 
By dynamically rebalancing adapters across GPUs and leveraging GPU Direct RDMA for remote access, \sysname maximizes throughput and minimizes tail latency under real-world workload drift. Evaluations on production traces from \orgname show that \sysname elicits up to 2$\times$ higher throughput, up to 9$\times$ lower TTFT, while using up to 50\% fewer GPUs under SLO constraints compared to state-of-the-art systems.

\end{abstract}

\section{Introduction}

Large language models (LLMs) are reshaping modern applications. Models like GPT \cite{gpt} and Llama\cite{llama}, pre-trained on vast corpora, exhibit strong general-purpose capabilities. These pre-trained (base) LLMs are increasingly fine-tuned for specific domains to optimize particular use cases, for instance, adapting Llama-2 for improved code generation\cite{codellama}. Emerging LLM platforms \cite{openai-ft, anyscale-ft, aws-ft, azure-ft} offer fine-tuning APIs and services that let developers customize models and build domain-specific applications. For example, OpenAI provides APIs to fine-tune GPT-4 and a Completions API to serve those fine-tuned models \cite{openai-ft}.

Low-Rank Adaptation (LoRA) \cite{lora, qlora} has emerged as a widely used parameter-efficient fine-tuning method for LLMs \cite{peft}. Instead of updating all model weights, LoRA learns low-dimensional updates captured by pairs of small matrices (called LoRA adapters) while the base model remains frozen. Practically, this method of LLM fine-tuning involves training only a LoRA adapter for the target domain, thereby significantly reducing the parameter count and compute requirement. 
For example, compared to fully fine-tuning GPT-3 (175B), LoRA reduces trainable parameters by over two orders of magnitude and lowers GPU usage by approximately $3\times$, while maintaining comparable model quality \cite{lora}.



\begin{figure}[h]
\centering
\vspace{-10pt}
    \centering   \includegraphics[width=0.99\linewidth]{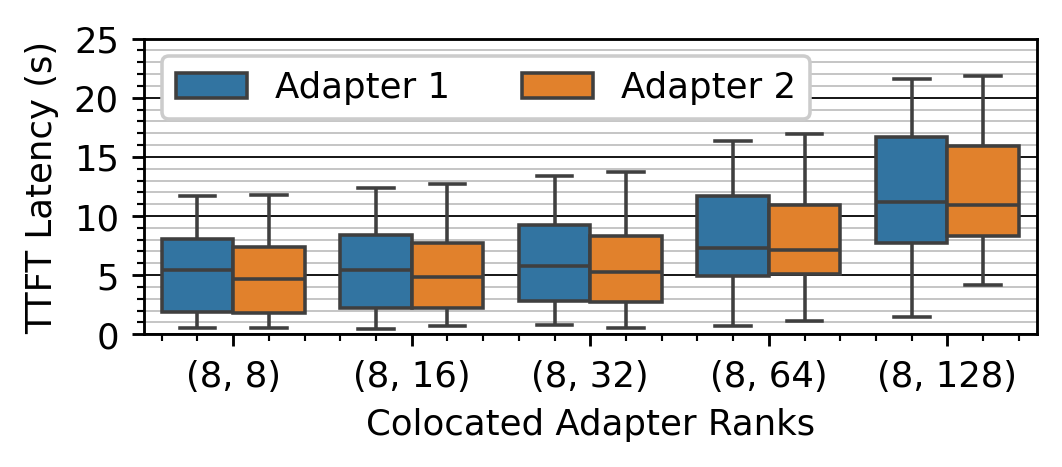} 
\caption{\textit{\textbf{Performance per adapter when two adapters are co-served from the same Llama 7B instance.} Higher rank heterogeneity leads to greater variability.}}
\label{fig:coloc}
\vspace{-12pt}
\end{figure}

While LoRA is cost-effective in training, its adapters introduce significant challenges at inference, specifically when multiple LoRA adapters need to be served off the same base model instance. Typical LoRA serving systems maintain thousands of LoRA adapters on inference servers, either in CPU or GPU memory, offering APIs for end users to access specialized LLM inference as per application requirements \cite{slora}. The inference server multiplexes different adapters to enable such multi-tenant LoRA serving using specialized CUDA kernels \cite{slora, punica} to co-batch requests to different LoRA adapters and execute them together. Consequently, adapters served on the same LLM instance experience performance interference; as shown in Fig \ref{fig:coloc}, requests to lower-rank adapters incur higher latency when co-served with larger-rank requests.



\begin{figure}[h]
\centering
    \centering
   \includegraphics[width=0.99\linewidth]{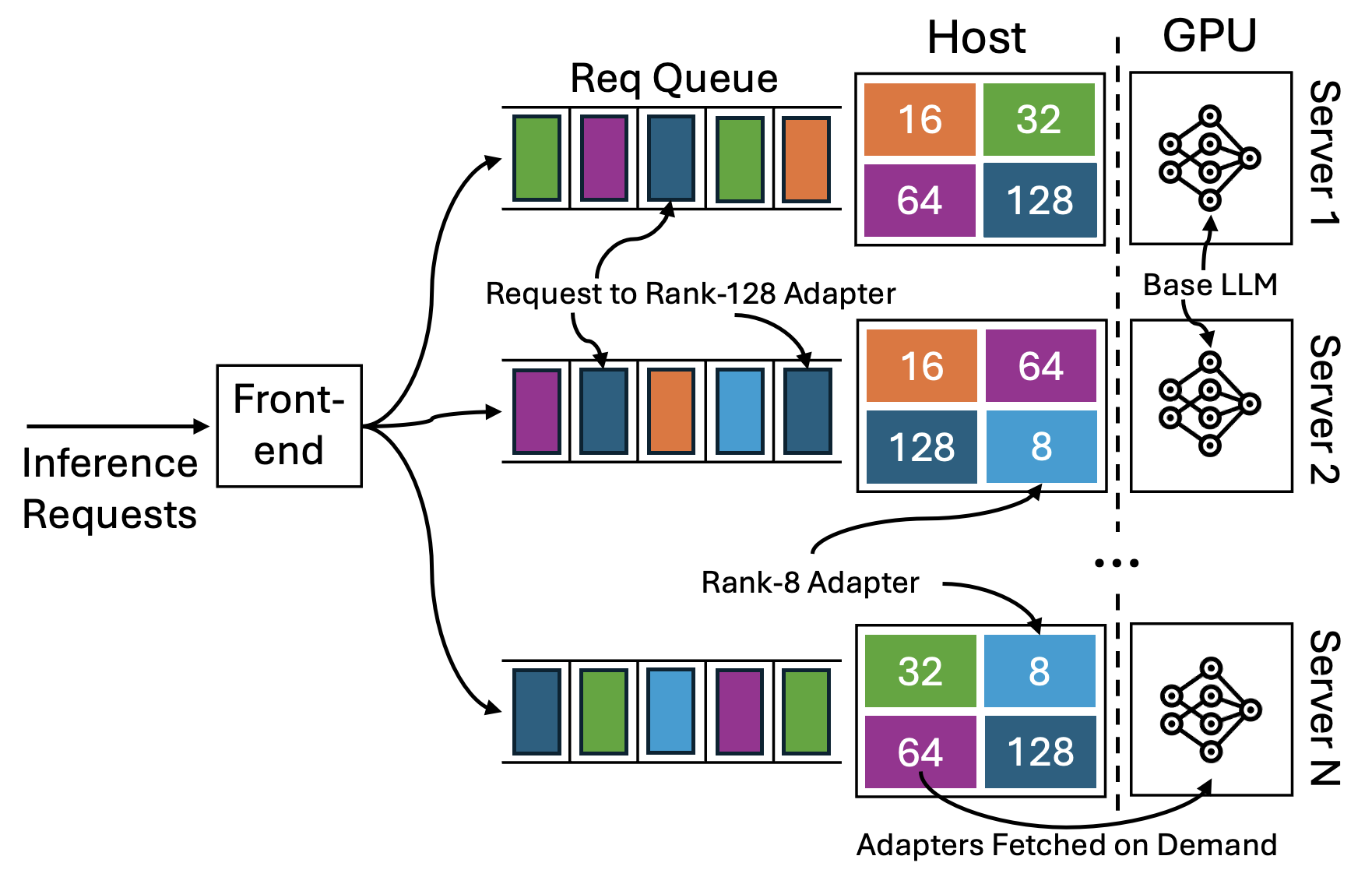} 
\caption{\textit{\textbf{A conventional LLM cluster serving LoRA workloads.}} 
}
\label{fig:convention}
\vspace{-6pt}
\end{figure}

Using these custom CUDA kernels \cite{slora, punica} along with hundreds of adapters on a cluster of LLM inference servers is the state-of-the-art and is used at \orgname and other cloud providers serving multiple LoRA adapters as shown in Fig \ref{fig:convention}. This entails two challenges, namely, \textit{managing the inherent scale and heterogeneity (adapter size, traffic etc.) of such workloads} and \textit{management of hundreds of LoRA adapters at a cluster level}.

\textit{Adapter and Workload Heterogeneity.} Each adapter has a different size (rank) and popularity (demand traffic). Prior work has shown that co-batching adapters of different ranks causes the smaller adapters to suffer in performance \cite{toppings}. Moreover, different adapters have different traffic (ingest rate, bursts etc.) depending on the application they serve, further compounding the heterogeneity. For example, Fig \ref{fig:coloc} shows that co-serving two adapters of rank-8 and rank-128 from the same Llama 7B LLM inference server causes the P95 TTFT (time-to-first-token) of the rank-8 requests to increase by 84\% when compared to a server serving the same amount of purely rank-8 requests.
 This latency increase forces systems to throttle request rates to meet SLOs (Service Level Objectives), reducing overall throughput and GPU efficiency. 
However, state of the art serving systems like vLLM \cite{vllm} and S-LoRA \cite{slora} remain oblivious to rank-induced heterogeneity, leading to performance degradation under real-workloads.

\textit{Managing Hundreds of LoRA Adapters.} \orgname uses hundreds of LoRA adapters to support a broad suite of services across hundreds of products used by millions of customers. Large cloud providers also store and serve their customers' LoRA Adapters as a service \cite{aws-sagemaker}. These are deployed using a cluster of LLM inference servers each running one base LLM instance and storing a subset of adapters in their CPU memory to be fetched on demand as shown in Fig \ref{fig:convention}. Each of these adapters see highly fluctuating request demand and statically placing a subset of adapters on each LLM server leads to load imbalance and SLO violations.
Another way is to replicate all adapters in the CPU memory of all servers. This is used in state-of-the-art \cite{toppings} to enable flexible request routing, but this significantly sacrifices CPU memory capacity on every server and may even be impossible at scale due to limited CPU memory. 
Considering an average general-purpose LLM has approximately 200B parameters \cite{llama-wikipedia} and is quantized in 8-bits \cite{8bit}, it {should} take up approximately 200GB of space. LoRA adapters are usually around 1\% of model size \cite{punica, lora} which comes to 2GB per adapter and 1TB for five hundred adapters in this case. 
Storing such high amount of data in every server's CPU memory is wasteful especially since  CPU memory is a precious resource for LLM inference scenarios involving long contexts \cite{longcontext} and retrieval-augmented-generation \cite{rag}.

\sysname addresses these performance degradation and memory pressure challenges by managing which adapters are served on which LLM inference server in the cluster. To do this, it takes into account adapter ranks and their request demands to ensure minimal heterogeneity within the adapters being served on a server. Moreover, it places locally only the adapters \textit{actually needed} by the server to alleviate memory pressure. 
The adapters are distributed across the cluster and they are fetched on demand when needed. 


Designing a system to address these issues 
efficiently has two technical challenges. The first is the determination of the optimal adapter placement, i.e., which adapter should be served on which (and how many) LLM servers. The second is keeping track of all adapters in the cluster and using this information when a necessary adapter is not present locally in a server. We propose a novel cluster orchestrator, \sysnamex, tailored for LoRA serving to address these two problems. 
To address the first problem, we propose a dynamic adapter placement technique that determines the ideal set of adapters to be placed on a particular LLM server taking into account the adapter rank and its request arrival history. This ensures load balancing, minimizes interference, and improves resource efficiency. To address the second problem, we propose simple indexing schemes to maintain a logical view of all the adapters in the cluster on every server and leverage GPUDirect RDMA over InfiniBand \cite{infiniband} interconnects to access them when necessary.


In summary, we make the following contributions:
{
\begin{itemize}
\item We analyze interference between adapters of diverse ranks, analyzing impact across multiple dimensions like deployment characteristics (e.g., parallelism), model size, and input size, and reveal implications of adapter rank heterogeneity on system throughput and SLO compliance.
\item Further, we characterize real-world LoRA serving workloads using production traces from \orgnamex, uncovering critical insights such as heavy-tailed request distributions, skewed adapter popularity across regions, and distinct arrival patterns. 
\item We design \sysnamex, a rank-aware, workload-adaptive framework for dynamic adapter placement and routing. \sysname dynamically rebalances adapters across LLM inference servers to mitigate rank-induced interference and optimize resource utilization under workload drift.
\item We evaluate \sysname on production workloads from \orgname and open source production traces \cite{azurepublicdataset}, demonstrating up to 2$\times$ higher throughput, up to 9$\times$ lower TTFT, while using up to 50\% fewer GPUs to meet SLOs compared to state-of-the-art systems. We will release our code and traces publicly after acceptance.
\end{itemize}}

\section{Background}
\subsection{Large Language Models}
\subsubsection{LLM Inference}
LLM inference generates output tokens from a given prompt in two stages: prefill and decode. In the prefill stage, the model reads the entire prompt, produces the first output token, and builds a key–value (KV) cache for each token. The decode stage then iteratively uses this KV cache to produce subsequent tokens, updating the cache as it generates each output token. Decoding stops once a termination condition is reached, typically when an end-of-sequence $(<eos>)$ token is generated.
\subsubsection{Parameter Efficient Fine-tuning}
Finetuning a model is a standard process which involves training a pre-trained model on specialized datasets to make them adapt to domain specific tasks.
Parameter-Efficient Fine-Tuning (PEFT) is a method which achieves comparable results by freezing most of the original parameters and training only a small subset of new or existing parameters \cite{peft}. This approach significantly reduces computational and memory requirements compared to full fine-tuning, making it more accessible and cost-effective. PEFT techniques like LoRA \cite{lora} and QLoRA \cite{qlora} allow models to achieve performance comparable to full fine-tuning for specific downstream tasks.
\subsubsection{Low-Rank Adaptation}
Low-Rank Adaptation fine-tunes a base model by inserting trainable low-rank matrices into its layers. The key driver of performance is the adapter size determined by the \textit{rank} of those matrices. Using a higher rank can capture richer updates and often improves accuracy. As tasks differ in complexity and their accuracy needs, application developers typically choose adapters with different ranks for different tasks while sharing the base model \cite{slora, lorapro}.

Given that adapters are typically $<1\%$ of their base model in size \cite{punica, lora}, this technique also reduces the memory requirements of online systems catering to diverse tasks. Moreover, recent works \cite{slora, punica} also enable executing multiple adapters in the same batch further improving throughput.

\vspace{-2mm}

\subsection{LoRA Serving Systems}
\subsubsection{Kernel Optimizations}
Research on kernel-level optimizations for LoRA adapter serving in LLM inference has largely focused on overcoming the batching and memory inefficiencies that arise when many requests share the same base model but request different adapters. Early LoRA serving approaches bound each request to a separate kernel invocation, which destroyed batch efficiency and led to GPU underutilization. Recent systems such as Punica \cite{punica} and S-LoRA \cite{slora} address this by redesigning the compute path. Punica introduces a segmented-gather GEMV (SGMV) kernel that fuses heterogeneous LoRA deltas into a single matmul, allowing the GPU to process many different adapters in one batched operation. This makes multi-tenant LoRA serving compute-efficient even when adapters and adapter ranks differ. 

\subsubsection{System Optimizations}
S-LoRA \cite{slora} pairs custom heterogeneous LoRA kernels with unified paging, a GPU-aware memory virtualization layer that stores a large number of adapters in CPU memory and pages only small active slices to the GPU alongside KV-cache blocks. This unified allocator reduces fragmentation and enables thousands of adapters to coexist without increasing GPU memory footprint. 
Toppings \cite{toppings} simultaneously uses CPUs to compute the low-rank adaptation for prefilling as the requested LoRA adapter is being loaded onto GPUs, to improve the TTFT. It also proposes a  rank-aware scheduling policy for load balancing requests but assumes that all adapters are replicated on each server in the cluster. dLoRA \cite{dlora} improves inference efficiency by dynamically merging adapters with the base model and migrating requests and adapters between different worker replicas. Chameleon \cite{chameleon} introduces adapter caching in GPU memory to prevent unnecessary loading of adapters from system memory. None of these systems take adapter rank heterogeneity into consideration while making scheduling decisions, leading to performance degradation due to interference between requests of different ranks.


\section{Motivation}
\label{sec:motivation}
\subsection{Performance}
In this section, we quantify the interference between adapters of diverse ranks, analyzing its impact across multiple dimensions, including parallelism, model size, and input size, and reveal implications of adapter rank heterogeneity on system throughput and SLO compliance.

\subsubsection{Input Sizes}
\begin{figure}[h]
\centering
\vspace{-10pt}
    \centering
   \includegraphics[width=0.99\linewidth]{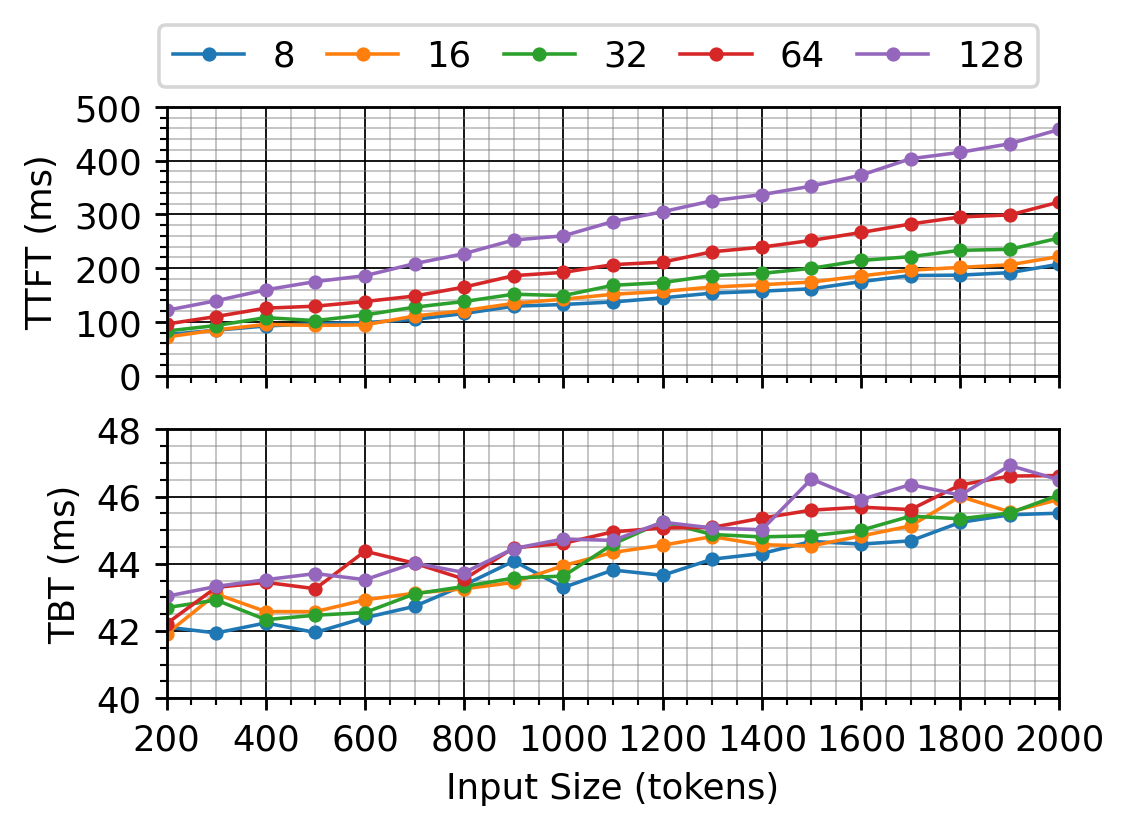} 
\caption{\textit{\textbf{Time-To-First-Token (TTFT) (top) and Time-Between-Tokens (TBT) (bottom) of different ranks against Input Size for Llama 7B.}}}
\label{fig:llama7b-tbt-ttft}
\vspace{-12pt}
\end{figure}

Serving adapters of larger ranks involves loading larger chunks from memory and computation on larger matrices, which naturally makes it slower than serving smaller ranks.  Fig \ref{fig:llama7b-tbt-ttft} shows these trends for a single request running in isolation. We observe that a rank-128 adapter takes $2.7\times$ the prefill time of a rank-8 adapter to process a prompt of size 2000. For decodes, the effect is subtle as the operation is largely memory bound. However, we still see a steady drop in performance with increase in input size. As input size grows, the performance impact of adapter rank becomes more pronounced because LoRA is applied to the Q, K, V, and O projection layers, whose compute cost grows with sequence length \cite{qlora}.

\subsubsection{Model Sizes}
\begin{figure}[hb!]
\centering
\vspace{-10pt}
    \centering
   \includegraphics[width=0.99\linewidth]{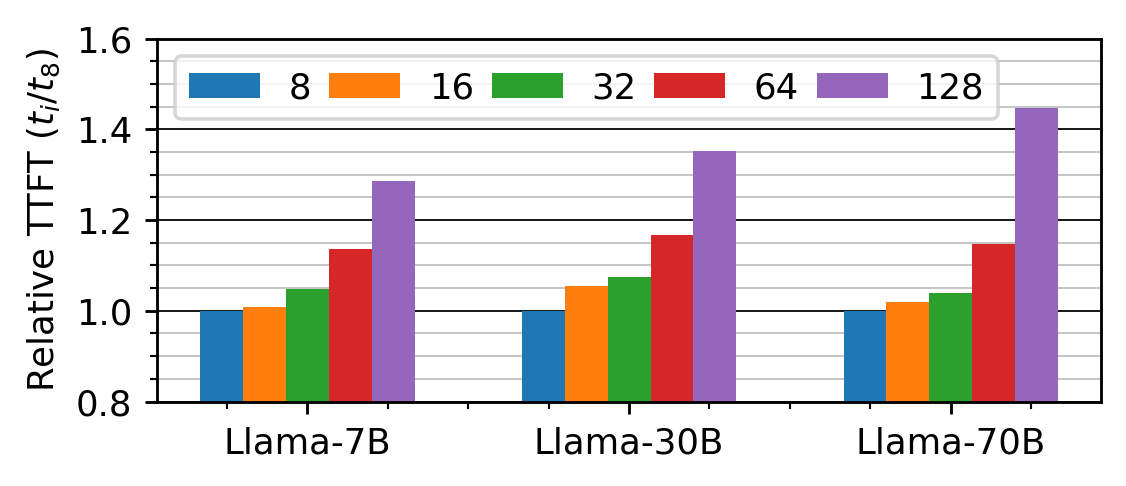} 
\caption{\textit{\textbf{Relative TTFT of different ranks on increasing model size.} Input size = 2000 and TP = 8.}}
\label{fig:model_size}
\vspace{-8pt}
\end{figure}

As model size increases, the impact of rank heterogeneity grows significantly, as illustrated in Fig \ref{fig:model_size}. Larger models introduce higher computational complexity, with wider hidden dimensions, deeper transformer stacks, and greater memory bandwidth demands. Consequently, LoRA’s low-rank matrices become larger and are injected into more projection layers, amplifying both compute and data-movement overhead. This increased pressure on the system magnifies the performance penalty from rank heterogeneity, reaching up to a 45\% degradation on Llama 70B.

\subsubsection{Tensor Parallelism} 
\begin{figure}[h]
\vspace{-10pt}
\centering
    \centering
   \includegraphics[width=0.99\linewidth]{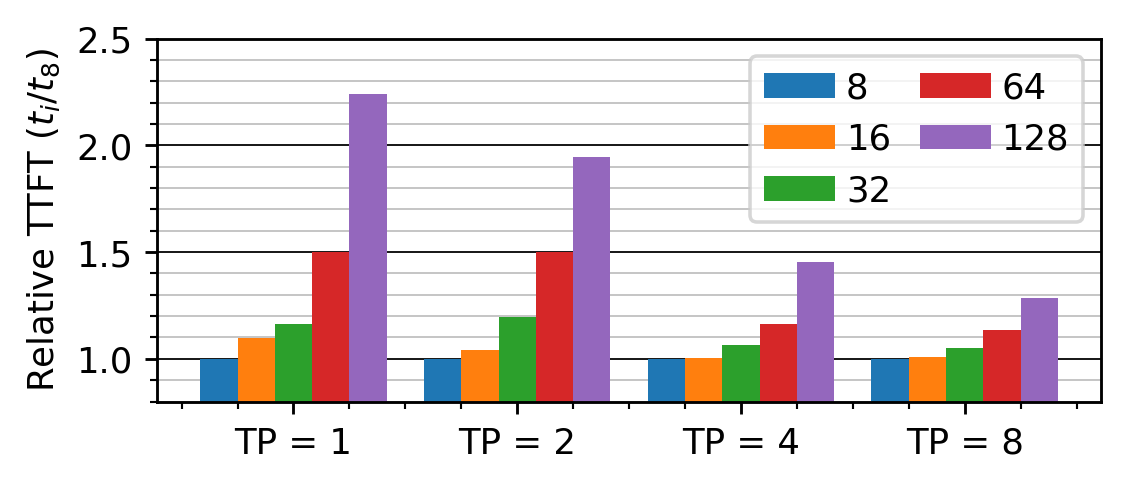} 
\caption{\textit{\textbf{Relative TTFT of different ranks on Llama 7B on increasing TP.} Input size = 2000.}}
\label{fig:tp}
\end{figure}

Production LLM deployments commonly rely on tensor parallelism (TP)\cite{prod-tp}. With increasing TP degree, the LoRA adapters are sharded across the GPUs\cite{slora} and hence the added computational cost of adaptation is divided across GPUs. Fig \ref{fig:tp} shows the performance impact of adapter ranks for different degrees of TP, where relative TTFT is computed by dividing the TTFT of each rank by that of rank-8. Though increasing TP diminishes the impact of rank heterogeneity, we still see a substantial 20\% increase in TTFT when using a rank-128 adapter with TP=8, on Llama 7B.

\subsubsection{Impact on Service Level Objectives (SLOs)}
\begin{figure}[h]
\centering
\vspace{-10pt}
    \centering
   \includegraphics[width=0.99\linewidth]{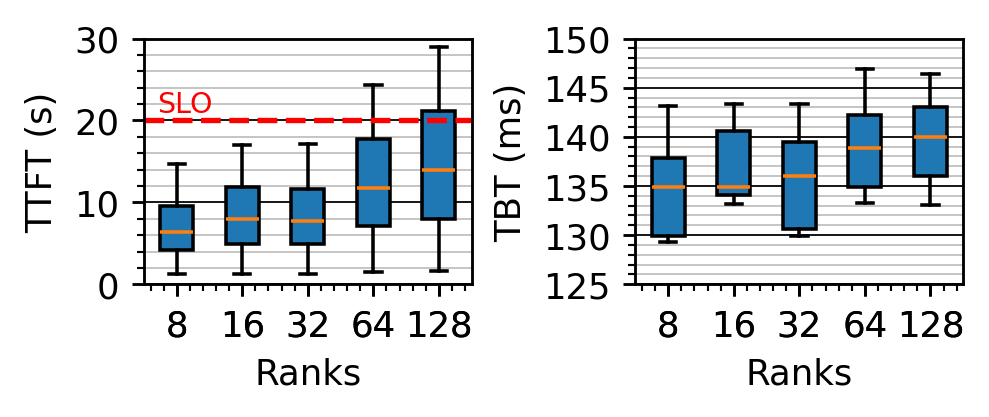} 
\caption{\textit{\textbf{4 RPS Poisson workload on Llama 7B with varying adapter ranks and fixed input (512) and output (128) lengths.}}}
\label{fig:tps}
\vspace{-10pt}
\end{figure}

Cloud providers operate under strict SLOs for tail latencies. For example, a common SLO for LLM Inference is that the P95 TTFT must be under 10s \cite{sageserve}. A direct consequence of varying performance with adapter ranks is that the same hardware can process fewer tokens per unit time when operating on higher ranks. Thus, to operate under SLO, more resources must be dedicated for larger ranks making their cost per token greater. Fig \ref{fig:tps} shows the performance of a simple 4 RPS (requests per second) Poisson arrival workload on different ranks using the same underlying hardware. Assuming a P95 TTFT SLO of 20s, Fig \ref{fig:tps} shows that high ranks like 64 and 128 violate SLO for the same workload whereas others do not.

\subsubsection{Effect of Co-serving Adapters of Different Ranks}
The kernels used for multi-tenant LoRA serving \cite{slora, punica} enable co-batching requests from different LoRA adapters. However, both kernels size their compute tiles and MMA pipelines to the maximum LoRA rank in the batch. In Punica \cite{punica}, the BGMV forces all requests, regardless of their actual rank, to execute GEMMs padded to the largest rank, inflating register usage, shared-memory requirements, and tile dimensions. On the other hand, S-LoRA's MBGMV \cite{slora} mitigates some padding overhead, but still inherits the same dependency, i.e., the fused kernel’s scheduling and memory layout are dictated by the highest rank present. As a result, the presence of high-rank requests in the batch causes low-rank requests to slow down, despite needing far fewer FLOPs.
Kernel-level profiling confirms that throughput and latency track the maximum rank \cite{toppings}.
Consequently, co-batching heterogeneous ranks in these kernels bottlenecks performance, causing smaller-rank adapters to pay the computational cost of the largest one.

Fig \ref{fig:coloc} depicts the prefill performance when two different adapters are co-served on a single host. Evidently, co-serving rank 8 with rank 128 makes the smaller rank's tail performance slower by 84\%. Moreover, greater rank heterogeneity leads to higher variability in performances, as indicated by the larger ranges in the box plot. This calls for improvements at the system level to alleviate these issues.

\subsection{Workloads}
\noindent 
We analyze production traces from LoRA deployments across regions for the week of October 22–29, 2025 at \orgname. The characterization provides a comprehensive view of adapter distribution, memory footprint, traffic patterns, and arrival trends, offering critical insights into workload heterogeneity and its implications for system design and resource allocation.
\noindent 
\subsubsection{LoRA in Production} 
\begin{figure}[h]
\centering
\vspace{-10pt}
    \centering
   \includegraphics[width=0.99\linewidth]
   {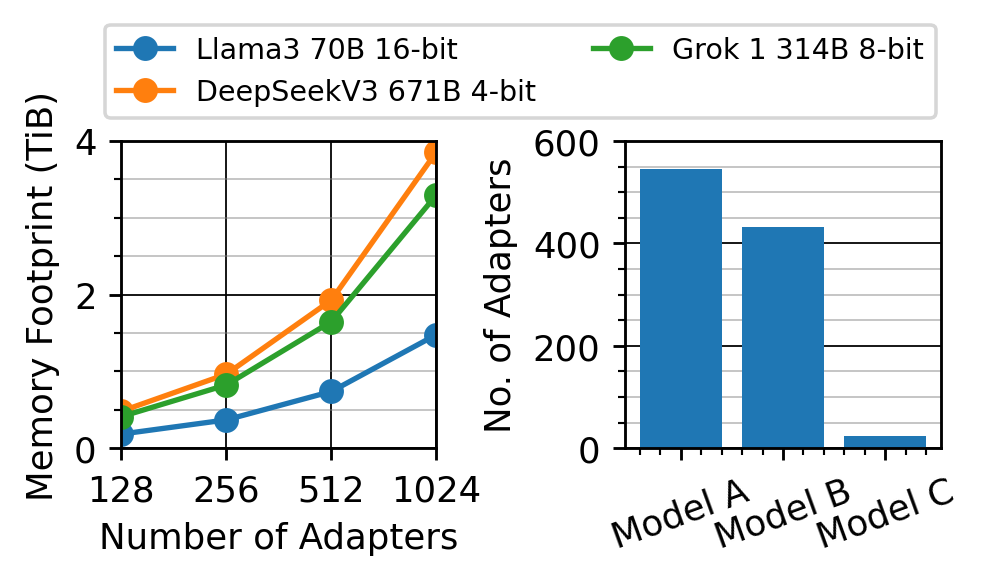}
      
\caption{\textit{\textbf{Estimated memory footprint of adapters (left) and no. of LoRA adapters for three base models at \orgname (right).}} \textit{Popular base models host hundreds of LoRA adapters with substantial memory footprints, making full colocation infeasible and motivating selective placement strategies.}}
\label{fig:adapterpopularity}
\vspace{-10pt}
\end{figure}

\begin{figure}[h]
\centering
\vspace{-10pt}
    \centering
   \includegraphics[width=0.99\linewidth]{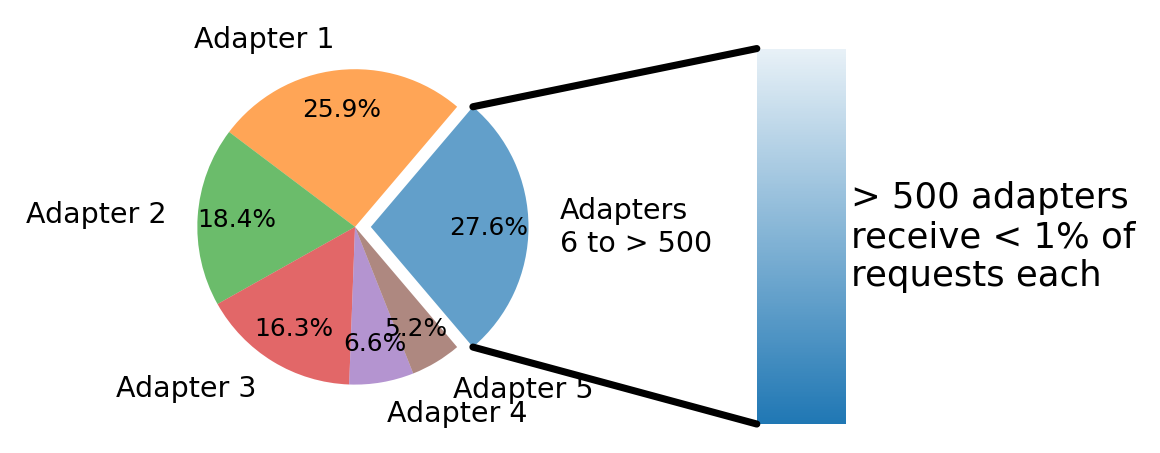} 
\caption{\textit{\textbf{Each adapter's request share for Model A at \orgnamex.} Top 5 adapters  account for more than 70\%  of requests, highlighting the need for demand-aware capacity allocation and co-location of low-demand adapters.}}
\label{fig:heavytail}
\vspace{-8pt}
\end{figure}
Fig \ref{fig:adapterpopularity} shows the number of adapters for the three most popular base LLMs deployed at \orgnamex. Certain base models host significantly more adapters than others, making it impractical to place all adapters on every LLM inference server \cite{toppings}, especially due to high memory footprint as illustrated in Fig \ref{fig:adapterpopularity}. Moreover, out of the over thousand adapters currently deployed in production, merely the top 5 account for more than 70\% of traffic as illustrated in Fig \ref{fig:heavytail}. Hence, they can be merged into the base model for serving on dedicated LLM Inference servers due to high demand \cite{sageserve}. The remaining 27.6\% traffic is distributed across the rest of the adapter, each of which receive way less than 1\% of the total traffic. Dedicating exclusive LLM inference servers for these may lead to resource under-utilization due to highly fluctuating request demands. Thus, in production, each server hosts a subset of adapters.

\begin{figure}[h]
\centering
\vspace{-10pt}
    \centering
   \includegraphics[width=0.99\linewidth]{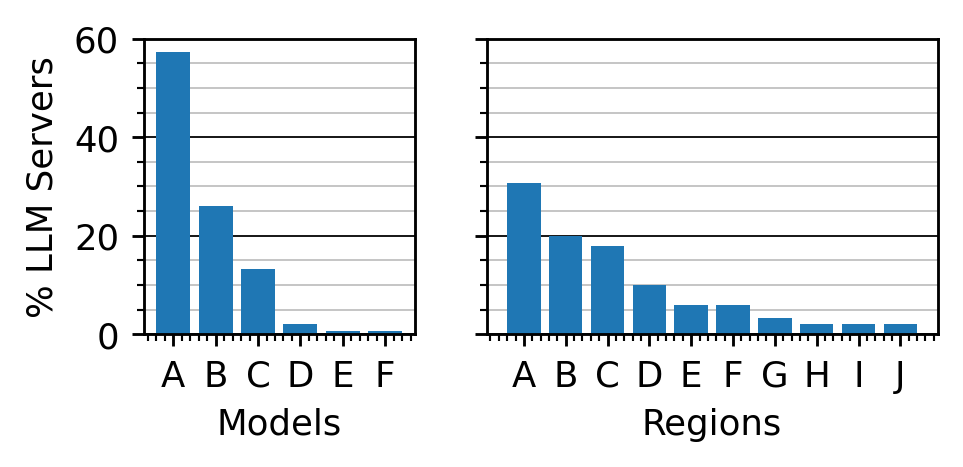} 
\caption{\textit{\textbf{Percentage of LLM servers used for each model (left) and at each region (right) at \orgnamex.} Resource consumption is concentrated in a few models and regions due to data boundary constraints, emphasizing that server-level optimization is more effective than global routing.}}
\label{fig:loradist}
\vspace{-8pt}
\end{figure}

As illustrated in Fig \ref{fig:loradist}, a few models dominate resource consumption, and this capacity is further concentrated in specific regions. Unlike vanilla LLM deployments, this imbalance arises because models are fine-tuned on customer data and are subject to data boundary constraints. Consequently, since capacity is regionally constrained rather than globally distributed, optimizing adapter placement at the server level offers greater potential for improving GPU utilization than global routing or load balancing strategies typically employed for vanilla LLMs.

\textit{Long-tail model popularity}:
Fig. \ref{fig:heavytail} shows the request distribution for each adapter over one week from Model A in Region A. The top five adapters account for more than 70\% of all requests.
This implies that tailoring capacity allocation to adapter-specific demand patterns enables additional GPU savings.
Co-locating multiple low-demand adapters on the same GPU, both spatially and temporally, can significantly improve resource utilization.

    

\begin{figure}[h]
\centering
\vspace{-10pt}
    \centering
   \includegraphics[width=0.99\linewidth]{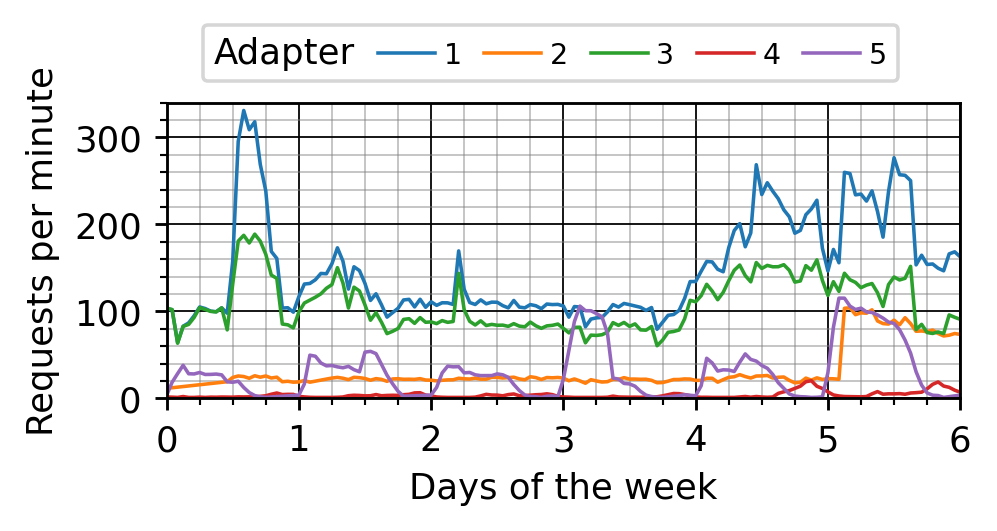} 
\caption{\textit{\textbf{Requests per minute for five adapters over one week in October 2025.} Arrival patterns of top adapters vary significantly, with drifts over time, highlighting the importance of workload-aware strategies for dynamic resource allocation.}}
\label{fig:rpm-week}
\vspace{-10pt}
\end{figure}

\textit{Arrival Patterns of Top 5 Adapters}:
Next, we examine the arrival patterns of the top five adapters for Model A in Region A. As shown in Fig \ref{fig:rpm-week}, each adapter exhibits a distinct trend in request arrivals over the observed week, captured through the moving average of requests per minute 
Most adapters display smooth activity with gradual drifts, though anomalies are present. Notably, Adapter 1 and 3 show varying load trends while adapter 5 follows diurnal patterns. Adapter 4 exhibits relatively stable demand whereas adapter 4 remains stable initially but suddenly sees a load surge towards the end of the trace. These fluctuating patterns highlight the potential benefits of dynamic rebalancing of resources allocated to adapters. This can improve the efficiency of capacity allocation, thereby improving GPU utilization and reducing resource overhead compared to static allocation strategies.

\subsubsection{Implications for System Design}
The presence of a few adapters that make up the majority of traffic conveys that GPU resource allocation should be based on adapter demand and not a simple uniform distribution across all adapters. The large number of adapters with little traffic must be efficiently multiplexed for effective resource utilization. 
This multiplexing cannot be static, since we see varying load patterns across adapters and across time, necessitating dynamic adapter placement based on adapter demand.



\section{\sysname Design}

Based on insights from our characterization (Section \ref{sec:motivation}), we design \sysnamex, an LLM Inference cluster orchestrator designed for serving LoRA adapters at scale on a pool of LLM Instances. \sysname  is designed to address the unique challenges faced by enterprises like \orgname while scaling heterogeneous LoRA adapter serving. Broadly, \sysname addresses two challenges, i.e., (1) \textit{the overheads of co-serving heterogeneous adapters} and (2) \textit{the memory capacity pressure of storing all adapters in each server}. 

\begin{figure}[h]
\centering
    \centering
   \includegraphics[width=0.99\linewidth]{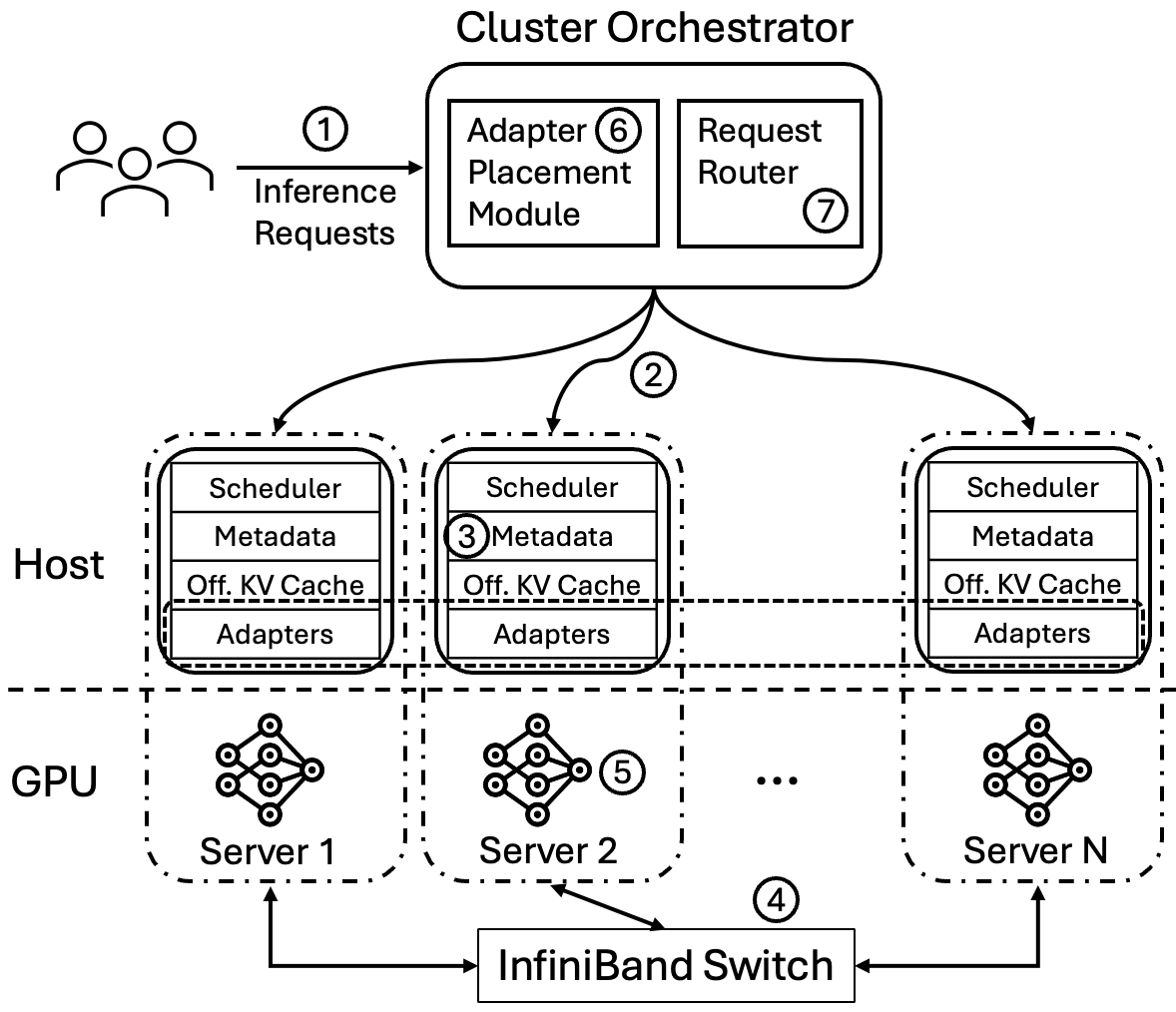} 
\vspace{2mm}
\caption{\textit{\textbf{\sysname architecture overview.} Each server stores and serves a subset of adapters managed by \sysnamex.}}
\label{fig:system}
\vspace{-6pt}
\end{figure}

To address the first issue, \sysname proposes a novel adapter placement and request routing algorithm that minimizes heterogeneity of adapters being served on a specific server while simultaneously balancing the request load evenly across the cluster by taking into account the popularity and workload demands of each adapter, at every time step. To accommodate thousands of adapters in a quickly accessible manner, \sysname only stores the adapters needed by a server in the current time step in its local host memory, ensuring that every adapter is present in at least one of the servers in the cluster. When the workload drifts and \sysname re-balances the adapter placement in the cluster, the new adapters are fetched over InfiniBand links using GPU-Direct RDMA on first access and then stored locally. The placement and routing module and the adapter migration module work synergistically, maximizing the system’s throughput and balancing the request load in the system.

\textbf{Architecture Overview.} Fig \ref{fig:system} shows an overview of the \sysname architecture. For sake of explanation, we assume that one LLM inference server runs one instance only. \sysname maintains a routing table in the cluster orchestrator that contains tuples of  $(adapter\_id, server\_id, \phi)$ for every adapter. The adapter corresponding to $adapter\_id$ is stored on all  servers corresponding to $server\_id$ across such tuples. An inference request \circled{1} specifying a particular adapter is received at the cluster orchestrator and routed \circled{2} to $server\_id$ with probability $\phi$. It is ensured that for every $adapter\_id$, $\Sigma \phi = 1$, implying the request is guaranteed to be executed on one of the servers where its adapter is allocated. In the host, \sysname looks up the adapter metadata to identify where the adapter is present \circled{3}. If the adapter is not in the local memory of the host, \sysname initiates a transfer over InfiniBand using GPUDirect RDMA to fetch it from the remote server where it is present \circled{4}. Finally, the LLM instance runs the inference \circled{5} and returns the result.
At the cluster orchestrator level, the \sysname service keeps track of how many requests each adapter receives. At every time step (configurable by the cluster admin), \sysname estimates the TPS (tokens per second) demand of every adapter and uses it, along with information regarding the rank and current location, to generate a balanced placement of adapters \circled{6}. Subsequently, the new mappings are updated in the request router \circled{7} and used for subsequent requests. 

\subsection{\sysname Adapter Placement}
\label{sec:place}
The adapter placement algorithm takes into consideration the load expected for every adapter in the cluster and makes allocation decisions to determine which server to use to serve each adapter. Allocation of an adapter to a server in \sysname is a function of the projected workload demand of that adapter, its rank and the target utilization of the cluster. By taking these factors into account, \sysname is able to generate adapter placements that minimize heterogeneity and balance request load across the cluster. For every \textit{adapter} with projected load $\ell$, the expression below shows the output of the algorithm where $s_1, s_2 \dots s_k$ are the $k$ servers in the cluster and $\phi_1, \phi_2 \dots \phi_k$ denote the fractional load (probability of serving) allocated to each server where $\Sigma \phi = 1$. 
\begin{equation}
    f(adapter, \ell) = \bigl[(s_1, \phi_1), (s_2, \phi_2), \dots , (s_k,\phi_k)\bigr]
    \nonumber
\end{equation}
\vspace{-10pt}
\begin{algorithm}
\caption{\sysname Algorithm}
\label{alg:adapter-placement}
\begin{algorithmic}[1]
\Function{assign\sysname}{servers, adapters, requestHistory, operatingPoints, prevAssignment}
    \State $\text{demandTPS} \gets \{\}$ \Comment{Step 1}
    \For{\textbf{each} adapter in adapters}
        \State $\text{adapterTPS} \gets$ \Call{getPrevTimeStepTPS}{adapter, requestHistory}
        \State $\text{demandTPS[adapter]} \gets$ \Call{\mbox{extrapolate}}{TPSHistory[adapter], adapterTPS}
    \EndFor
    \State $\text{targetUtil} \gets 0$
    \State $\text{rankUtil} \gets \{\}$
    \For{\textbf{each} rank in \Call{uniqueRanks}{adapters}}
        \State $\text{rankUtil} \gets$\\$  \Sigma_{rank}\text{demandTPS} / \text{operatingPoints[rank]}$
        \State $\text{targetUtil} \gets  \text{targetUtil} + \text{rankUtil[rank]}$
    \EndFor
    \State $\text{targetUtil} \gets \text{targetUtil}/\Call{len}{\text{servers}}$

    \State $\text{rankServerBudget} \gets \{\}$ \Comment{Step 2}
    \For{\textbf{each} rank in \Call{uniqueRanks}{adapters}}
        \State $\text{rankServerBudget[rank]} \gets$ \Call{round}{rankUtil[rank]/targetUtil}
    \EndFor

    \State $\text{assignment} \gets \{\}$ \Comment{Step 3}
    \For{\textbf{each} rank in rankServerBudget} \\
        \Call{fractionalBinPacking}{adapters, rankServerBudget[rank], targetUtil, assignment}
    \EndFor

    \State $\text{leftovers} \gets$ {\scshape getUnassignedAdapters}(adapters, assignment) \Comment{Step 4}

    \Call{sortDescendingRank}{leftovers}
    \For{\textbf{each} adapter in leftovers}
    
        \Call{allocateHighestMaxRank}{adapter, assignment} \Comment{Allocates the adapter to the server with highest max rank and least utilization}
    \EndFor

    \Call{permuteAssignment}{assignment, \mbox{prevAssignment}}\Comment{Step 5}
    
    \Call{updateRoutingTable}{assignment}\Comment{Step 6}
    \For{\textbf{each} server in servers}

    \Call{updateAdapterMapping}{assignment, server}
    \EndFor

    
\EndFunction

\end{algorithmic}
\end{algorithm}

We profile the servers a priori, to estimate the operating point of each rank under SLO constraints, i.e., the maximum number of tokens per second the LLM inference server can process using an adapter of a specific rank under SLO constraints. Algorithm \ref{alg:adapter-placement} illustrates \sysnamex's adapter placement logic, using necessary abstractions for brevity. The algorithm follows the following basic steps at every time step:
\begin{enumerate}
    \item Estimate the TPS demand per adapter and calculate average target utilization per server using operating point information.
    \item Calculate the server budget per rank, i.e., the number of servers that can be dedicated for that rank.
    \item Pack the adapters of ranks which have non-zero server budget assigned using fractional bin packing.
    \item Preferentially allocate each of the remaining adapters on a server with higher maximum rank, if possible.
    \item Permute the placement across servers to ensure minimal deviation with previous placement.
    \item Update routing table and local adapter metadata.
\end{enumerate}

\begin{figure}[h]
\centering
    \centering
   \includegraphics[width=0.99\linewidth]{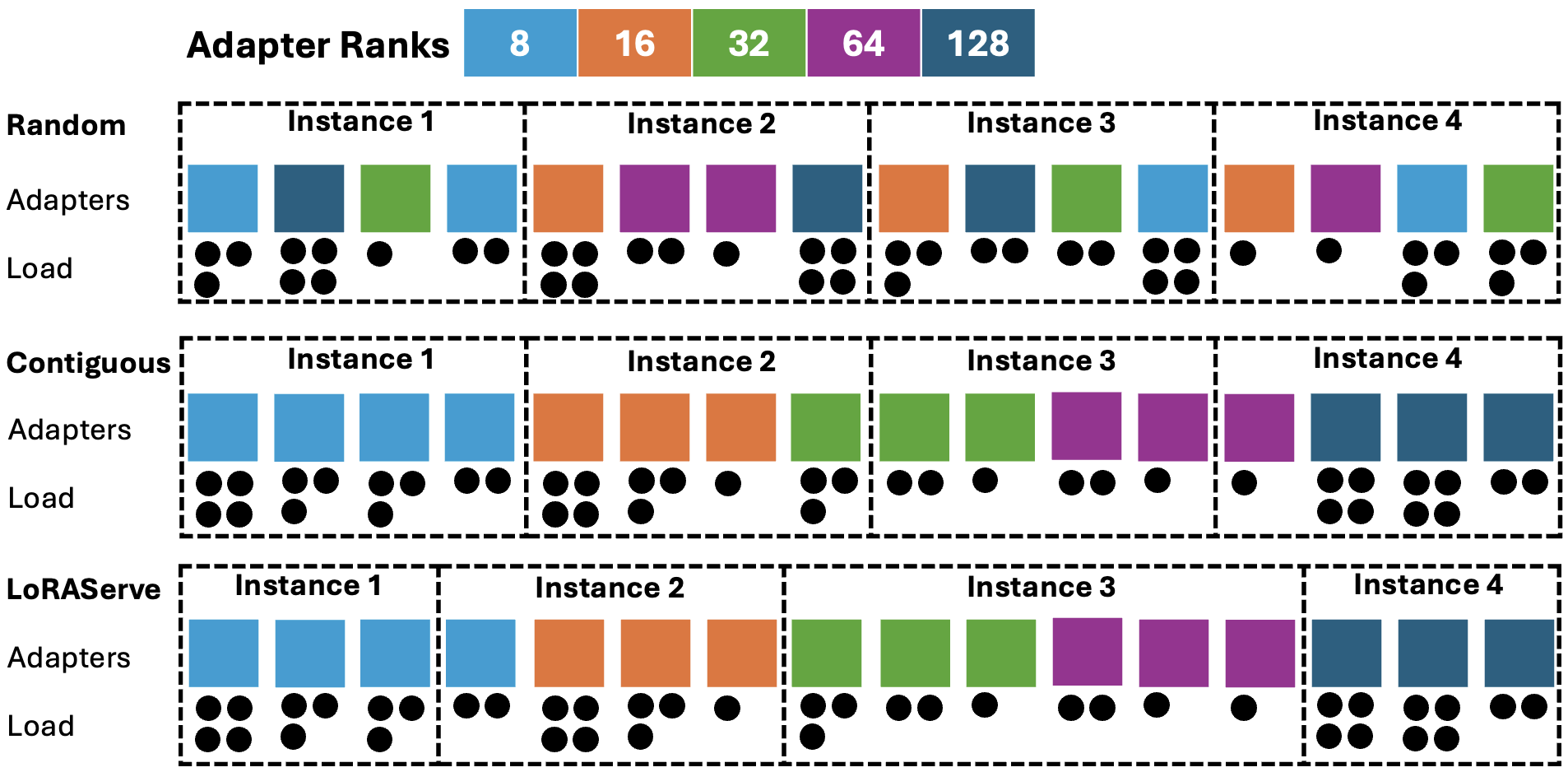} 
\vspace{-2mm}
\caption{\textit{\textbf{A high-level illustration of adapter placement by \sysname and baselines.}}}
\label{fig:intuition}
\end{figure}

Fig \ref{fig:intuition} shows a high level visualization of \sysnamex's adapter placement along with baselines. \textit{Random} placement does a good job of balancing the request load but does poorly in alleviating rank heterogeneity whereas naive rank-based \textit{Contiguous} placement does a good job of mitigating rank heterogeneity but is not able to balance load. This results in the former underutilizing resources and the latter having poor tail latency. \sysname on the other hand is able to optimize for both dimensions achieving better performance. 

\subsection{\sysname Distributed Adapter Pool}
The \sysname Adapter Placement (\ref{sec:place}) and routing mechanism ensures that an LLM instance is only responsible for processing requests corresponding to a relatively small subset of adapters for every time step duration. Moreover, it also ensures that the union of adapters assigned to every instance is the universal set of adapters that the cluster may need to use. The adapters assigned to a specific instance are stored in the main memory of the host. Combined, all locally stored adapters give the abstraction of the \textit{distributed adapter pool} in \sysnamex. An adapter may be assigned to one or more LLM servers depending on its popularity and demand.


The \sysname service running on the cluster orchestrator maintains an in-memory map of all the adapters in the cluster to help identify where they are stored. 
At every timestep, the adapter placement routine updates the placement and this map based on the latest adapter popularity patterns.

\begin{figure}[h]
\centering
    \centering
   \includegraphics[width=0.90\linewidth]{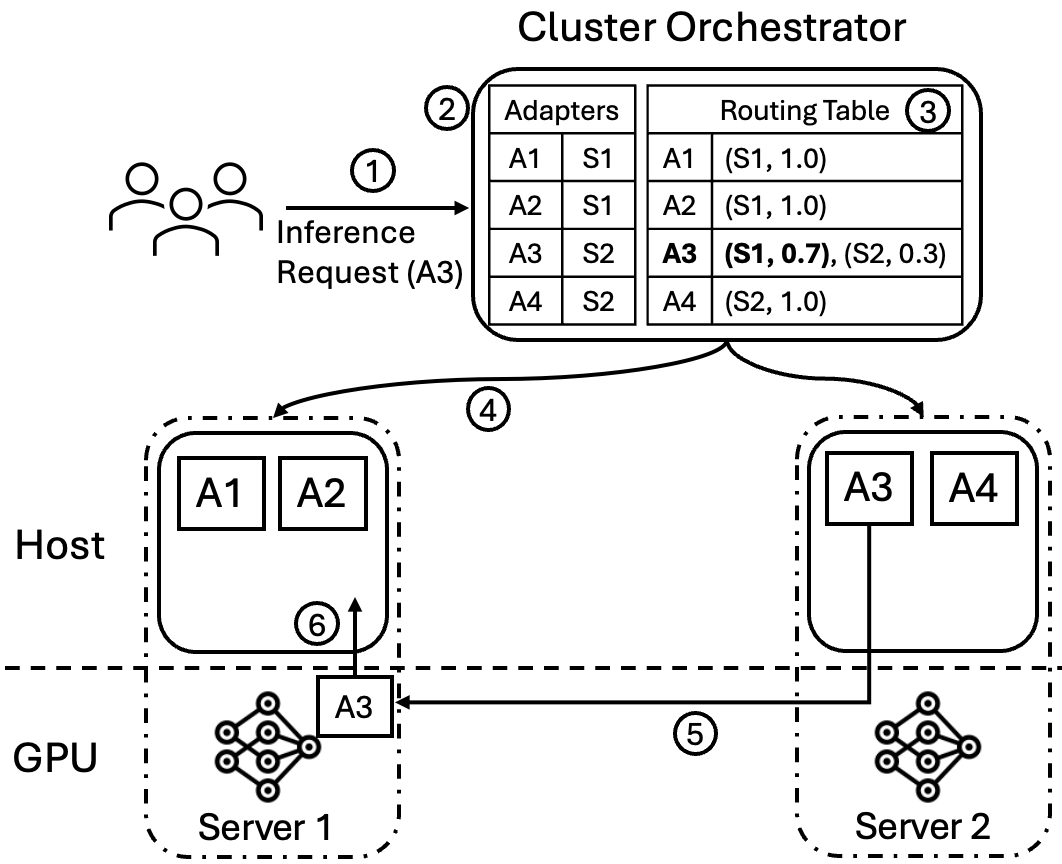} 
\vspace{1mm}
\caption{\textit{\textbf{\sysname Distributed Adapter Pool.} The cluster orchestrator maintains a table containing the location of each adapter in the cluster.}}
\label{fig:migration}
\vspace{-8pt}
\end{figure}

Consider the example illustrated in Fig \ref{fig:migration}. Assume that the adapter placement module recently changed the allocation of $A3$ from being served entirely on Server 2 ($A3: (S2, 1.0)$), to serving only 30\% of its traffic on Server 2 and the rest on Server 1 ($A3: (S1, 0.7), (S2, 0.3)$), but the actual migration hasn't occurred yet. When a request for adapter $A3$ arrives at the cluster orchestrator \circled{1}, \sysname looks up the adapters table \circled{2} to figure out where the adapter is present. Subsequently, it looks up the routing table \circled{3} and decides which server to send the request to. Assume that $S1$ is chosen for this example. The router then sends the adapter location information along with the inference request to $S1$ \circled{4}. Since the adapter is absent in $S1$, it is fetched from $S2$ over InfiniBand interconnects using GPUDirect-RDMA. To do so, the adapter must first be copied from host memory to GPU in $S2$ and then transferred to $S1$ \circled{5}. Once the request is processed, the adapter is saved in the host memory of $S1$ for future requests. Given that these operations are deterministic, the cluster orchestrator updates the adapter table to $A3: (S1, S2)$ upon successful completion of the request. In case if the adapter was no longer needed at $S2$ (i.e., the routing table had ($A3: (S1, 1.0)$)), $A3$ would be deleted from $S2$ after being copied to $S1$.

\begin{figure}[h]
\centering
    \centering
   \includegraphics[width=0.99\linewidth]{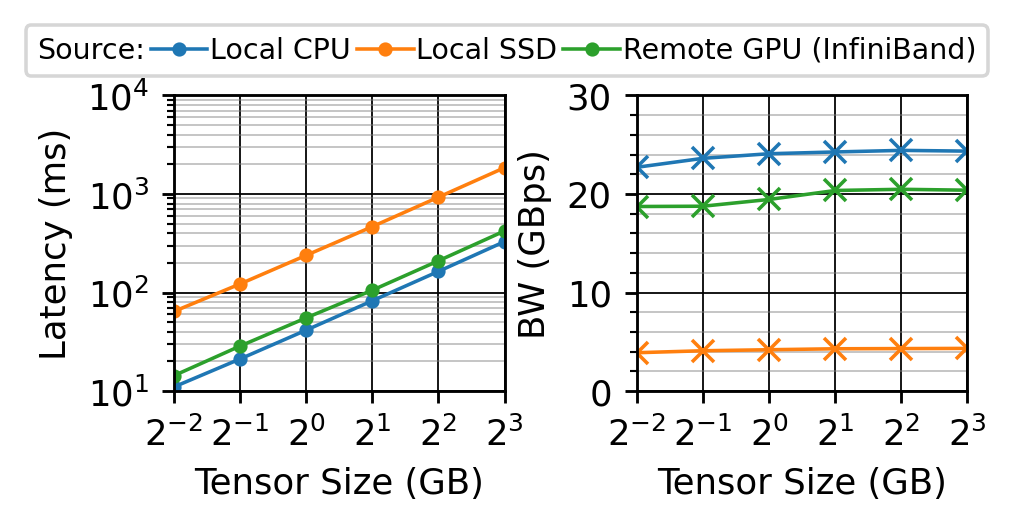} 
\caption{\textit{\textbf{Latency of fetching a tensor from different sources.}}}
\label{fig:interconnect}
\vspace{-5pt}
\end{figure}

Transferring adapters over InfiniBand \cite{infiniband} between GPUs of different servers incurs similar latency as transferring it over the same host's memory channel from local memory to GPU. Moreover, we explored the possibility of replicating all adapters in the local SSD of each server but the latency and bandwidth was found to be prohibitive. Fig \ref{fig:interconnect} benchmarks this trend. InfiniBand enabled us to perform the transfer operation without increasing the overheads.

\section{Evaluation}

We implement \sysname on top of S-LoRA \cite{slora}, a recent system built to serve thousands of LoRA adapters at scale. It is compatible with any LLM Inference framework that supports LoRA adapters and can also be integrated into common LLM cluster orchestrators \cite{llmd, vllm-prod-stack} used in production due to its modular design. 
\subsection{Hardware}
We run our experiments on Azure Machine Learning Compute Clusters \cite{aml}, one containing Standard\_ND96asr\_v4 VM\cite{azure-vm1} nodes and the other containing Standard\_NC24ads\_A100\_v4 VM\cite{azure-vm2} nodes. 
Each node in the former is equipped with 8$\times$ Nvidia A100 SXM 40GB GPUs connected through NVLINK 3.0, an AMD EPYC 7V12 CPU with 96 cores along with 900 GiB of host memory and 6TiB disk storage. Internode communication is through InfiniBand \cite{infiniband} and Azure Accelerated Networking \cite{azure_accl}. 
In the latter, each node contains 1$\times$ Nvidia A100 80GB GPU, an AMD EPYC 7V13 CPU with 24 cores, 220GiB of host memory and 960GiB of disk storage.
\subsection{Experiment Setup}
Due to limited resources, we run most of our experiments on a cluster of 4 LLM inference servers and validate scalability of \sysname on up to 12 servers. We use Llama 7B with TP = 4 on a uniform poisson trace of specified RPS, unless stated otherwise.

\subsection{Models}
We evaluate \sysname on the Llama family of models \cite{llama-wikipedia}. For majority of our experiments, we use Llama-7B. However, we also show \sysnamex's sensitivity to model sizes using Llama-7B along with Llama-30B and Llama-70B. While we present the results with Llama series of models, we also experimented with other models like OPT \cite{opt}. We observed similar performance trends.

\subsection{Baselines}
We compare \sysname against the following state-of-the-art baselines, keeping the underlying kernel the same across \sysname and the baselines for a fair comparison:
\begin{enumerate}
    \item \textit{S-LoRA Random} allocates adapters randomly to servers in the cluster, each running S-LoRA \cite{slora}. This adapter placement resembles the one used at \orgname.
    \item \textit{S-LoRA Contiguous} orders the adapters being served on the cluster by their ranks and places an equal number of adapters on each server contiguously. This ensures that ranks close to each other are co-located.
    \item \textit{Toppings} \cite{toppings} is a state of the art LoRA serving system that proposes a new load aware scheduler. The basic idea is to route the incoming request to the globally optimal server after considering the requests currently being served and queued at each server.
\end{enumerate}

\subsection{Traces}
\label{sec:traces}
\begin{figure}[h]
    \centering
    \vspace{-12pt}
        \centering
       \includegraphics[width=0.99\linewidth]{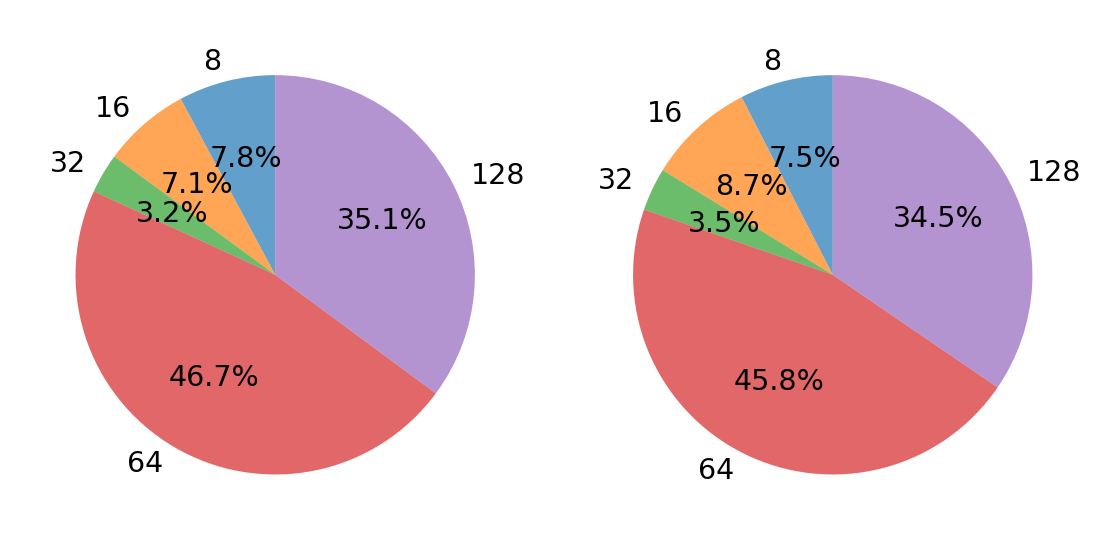} 
    \vspace{-6mm}
    \caption{\textit{\textbf{Rank-wise request (left) and token (right) distribution of production trace.}}
    }
    \label{fig:prod_trace}
    \end{figure}
We evaluate \sysname on production traces of real users interacting with \orgnamex's services that use LoRA adapters at scale. The trace contains 250,138 requests over 8 hours going to 5 adapters of different ranks used in production services as illustrated in Fig \ref{fig:prod_trace}. We divide each adapter's requests in the trace by annotating the traces with different adapter names of the same rank following a power law distribution for adapter counts within a rank, with $\alpha = 1$ totaling to 50, 100 and 200 to model three scenarios involving different total number of adapters. For evaluating on a desired RPS, we scale the timestamps proportionally to retain the original arrival pattern. The traces have the following attributes for every request: request\_id, model, adapter, prompt\_length, output\_length and timestamp.

We also evaluate \sysname on open-source Azure Public Dataset \cite{azurepublicdataset} traces from 2024. As these traces lack timestamps and adapter names, we annotate them with these attributes ourselves using some common assumptions used in prior work.
For timestamps, we assume the arrival pattern of the requests to be either uniformly distributed or following a Poisson process \cite{slora} as used in many prior works on LLM inference. We use a total of 25 adapters of ranks 8, 16, 32, 64 and 128 in our experiments as used by prior work \cite{toppings, chameleon}. For adapters' rank popularity, we evaluate on the following:
\begin{enumerate}
    \item Uniform: This assumes that all adapter rank popularities are uniformly distributed throughout the trace.
    \item Shifting Skew: This is illustrated in Fig \ref{fig:skew}. In the beginning, rank 128 gets half the traffic and the other half is distributed uniformly between the other four ranks. This skew shifts linearly until, at the end of the trace, rank 8 sees half the traffic and the rest is divided uniformly among the other ranks.
    \begin{figure}[h]
    \centering
        \centering
       \includegraphics[width=0.90\linewidth]{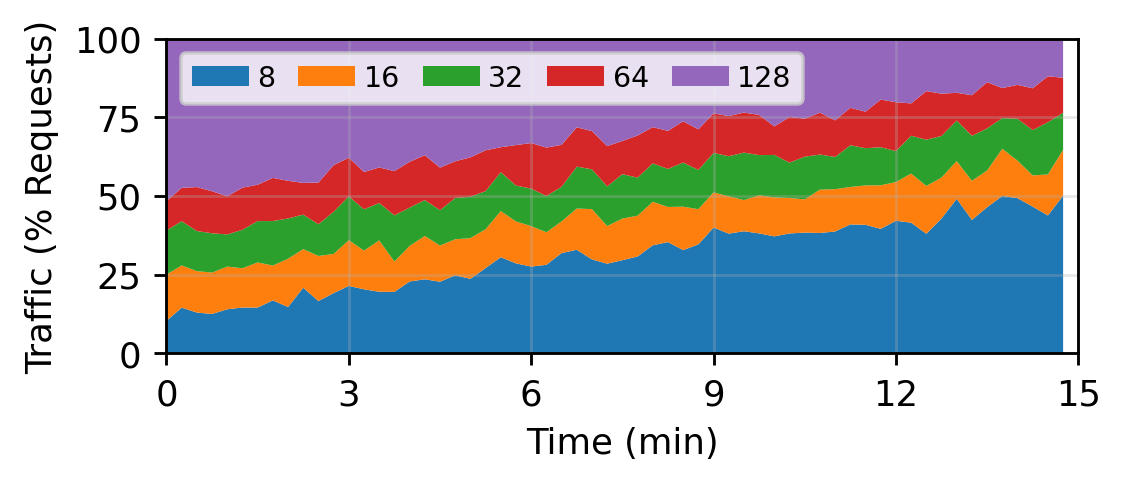} 
    \caption{\textit{\textbf{Shifting skew in adapter rank popularities.}}}
    \label{fig:skew}
    \vspace{-5pt}
    \end{figure}
    \item Exponential: This assumes that the popularities of ranks are exponentially distributed \cite{chameleon} with the smaller ranks being more popular as done by prior work.
\end{enumerate}

Combining both dimensions, i.e., request arrival pattern and adapter popularity, we get six unique traces that we use for subsequent evaluation.

\label{eval:perf}
\begin{figure}[h]
    \centering
    \vspace{-10pt}
        \centering
       \includegraphics[width=0.99\linewidth]{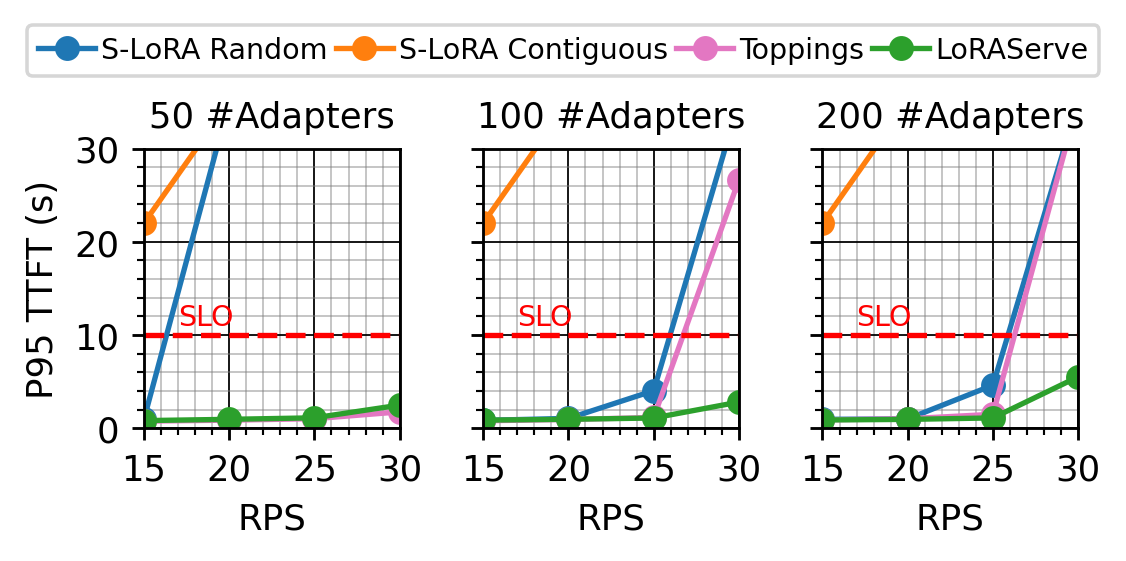} 
    \caption{\textit{\textbf{Evaluation on production traces from \orgname.}}}
    \label{fig:prod}
    \vspace{-5pt}
\end{figure}
\begin{figure}[h]
    \centering
    \vspace{-10pt}
        \centering
       \includegraphics[width=0.99\linewidth]{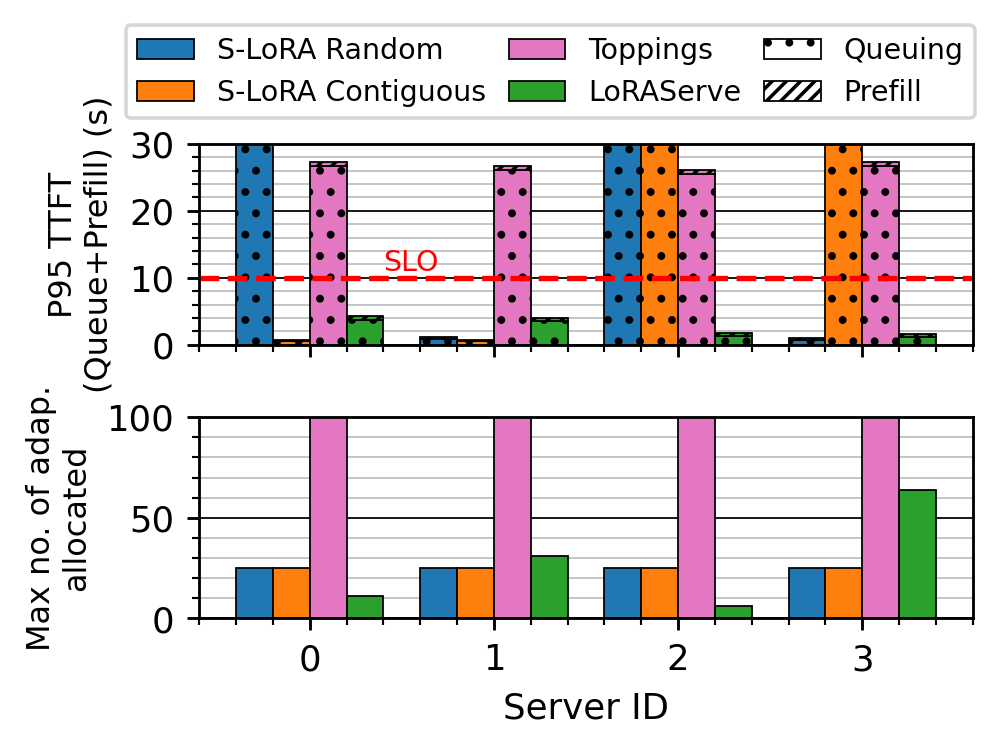} 
    \caption{\textit{\textbf{Server-wise queuing and prefill time (top) and maximum no. of adapters used (bottom)} for 30 RPS 100 \#Adapters experiment on production traces.}}
    \label{fig:queue}
    \vspace{-6pt}
\end{figure}
\subsection{Performance}
Fig \ref{fig:prod} shows the performance of \sysname and baselines on production traces from \orgnamex. As described earlier, we model three scenarios where we have a total of 50, 100 and 200 adapters each. We observe that \sysname is able to serve upto 20\% higher request throughput with respect to Toppings \cite{toppings} and upto 2$\times$ higher than S-LoRA \cite{slora}, under SLA. This translates into GPU savings of 17\% compared to Toppings and 50\% compared to S-LoRA when serving the same workload. This can be attributed to \sysnamex's rank aware adapter allocation for load balancing. Fig \ref{fig:queue} (top) shows the tail TTFT latency observed on each LLM inference server in the cluster. It is evident that S-LoRA Random and S-LoRA Contiguous, because of static allocation of adapters to servers, fail to balance the load evenly causing some servers to have excessively high load and time out. Toppings, on the other hand, performs load balancing at the request level, which is more granular, and does a much better job at having uniform performance on all servers. However, due to its rank agnostic behavior, it ends up sending high rank requests to all servers, causing interference leading to high amount of queuing and high TTFT. As \sysname is rank-aware, it is able to balance adapter placement dynamically and reduce queuing leading to upto $9\times$ lower TTFT. 
Furthermore, Fig \ref{fig:queue} (bottom) shows the maximum amount of adapters needed on every server by the four systems. We observe that \sysname reduces the amount of space needed for adapter storage by upto 16$\times$ with respect to state of the art Toppings. This is because \sysname allocates adapters to servers after taking into account both, the demand and rank of every adapter. Thus, a small number of same-ranked highly popular adapters get a dedicated server (server 0) and many sparsely used adapters get put together (server 3).

\begin{figure*}[htbp!]
\centering
    \centering
   \includegraphics[width=0.99\linewidth]{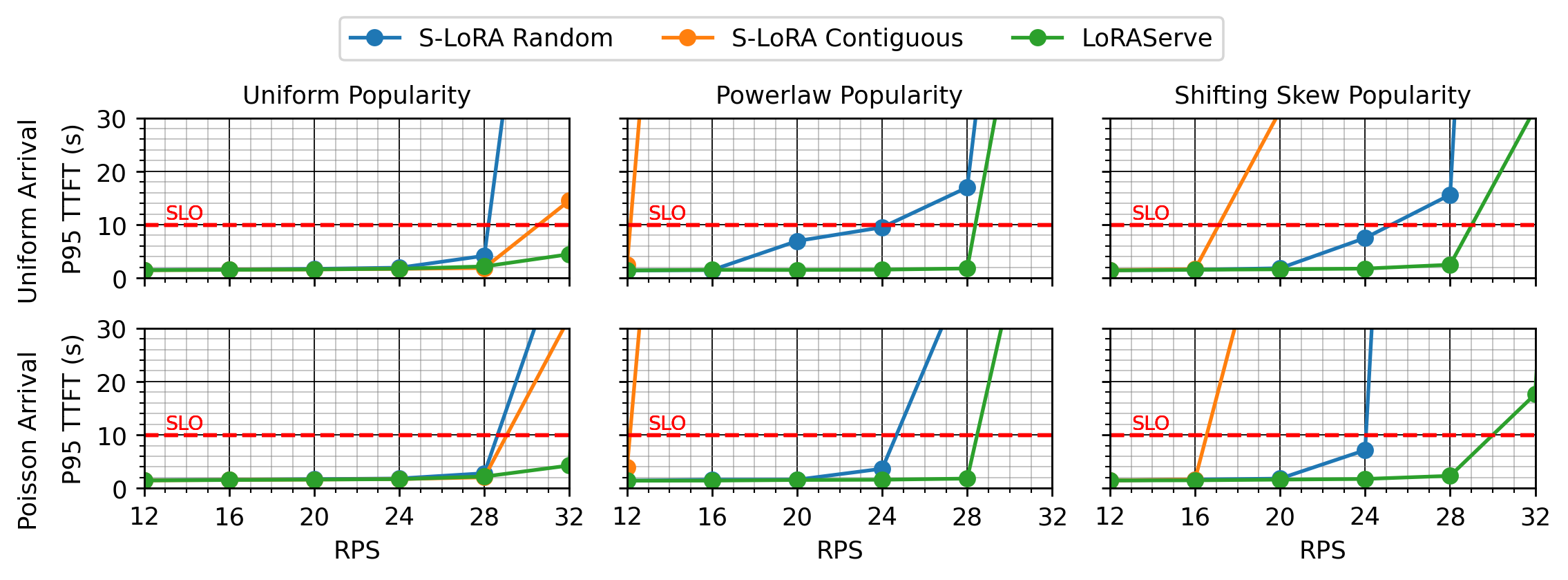} 
\caption{\textit{\textbf{TTFT performance on different workloads.} \sysname acheives up to 9$\times$ improvement in TTFT over baselines.}}
\label{fig:results-grid-ttft}
\vspace{-10pt}
\end{figure*}
\begin{figure*}[htbp!]
\centering
    \centering
   \includegraphics[width=0.99\linewidth]{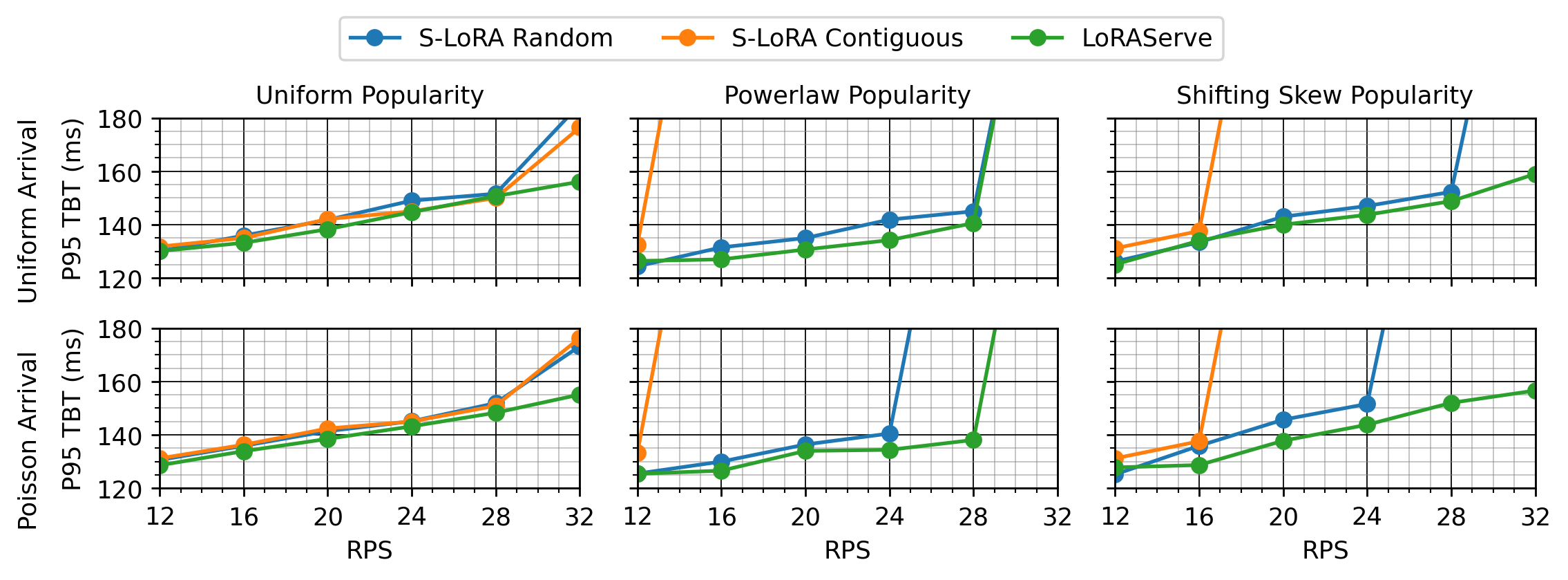} 
\caption{\textit{\textbf{TBT performance on different workloads.} TBT either remains similar in most cases or improves by up to 15\%.} 
}
\label{fig:results-grid-tbt}
\vspace{-10pt}
\end{figure*}
Fig \ref{fig:results-grid-ttft} and Fig \ref{fig:results-grid-tbt} show the performance of \sysname on the six traces derived from open-source production data (section \ref{sec:traces}). While \sysname beats the baselines in all cases, the general trend in Fig \ref{fig:results-grid-ttft} shows that S-LoRA Contiguous has reasonable performance when the popularity of adapters are uniformly distributed as it places approximately equal number of adapters having ranks close to one another on every server. This ensures good load balancing as every adapter receives similar traffic. However, it times out very quickly on other distributions. On the other hand, we observe S-LoRA Random to have passable performance in all cases except when the adapter demand changes (Shifting Skew) in the request trace. Moreover, S-LoRA Random also  times out at high request rates for most workloads and isn't able to scale at high traffic.

As \sysname accounts for both adapter rank and its demand, it is able to elicit much better performance and serve significantly higher request throughput on the same hardware. Quantitatively, we observe a TTFT improvement of upto $9\times$.
Furthermore, these improvements are achieved without any drop in request TBT (Fig \ref{fig:results-grid-tbt}). Instead, \sysname delivers up to a 15\% TBT improvement in some cases over the baselines due to better load balancing across servers in the cluster.

\subsection{Scalability}
\begin{figure}[h]
\centering
    \centering
\includegraphics[width=0.99\linewidth]{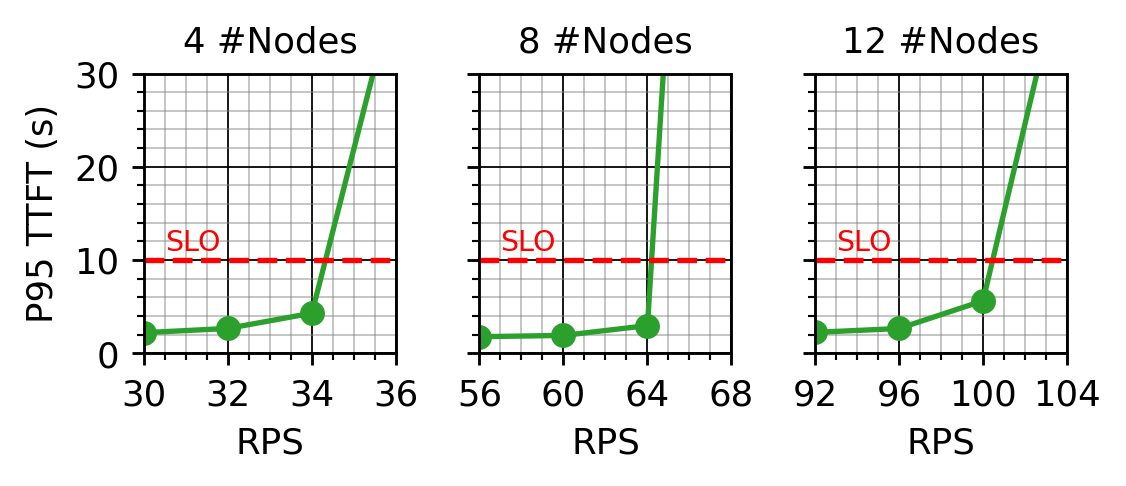} 
\caption{\textit{\textbf{Scalability of \sysname on up to 12 servers.}}}
\label{fig:scalability}
\end{figure}
Fig \ref{fig:scalability} shows the performance of \sysname when scaled to large number of nodes, i.e., clusters of size 8 and 12 LLM servers along with 4. We also scale the number of adapters and request traffic being served on the cluster proportionally with the number of nodes to evaluate weak scaling \cite{gustafson, aerialdb}. It can be observed that \sysname scales well with the number of nodes and is able to sustain a proportional amount of traffic with every added node. For example, in Fig \ref{fig:scalability}, \sysname is able to serve approximately 32 RPS workload within a 10s P95 TTFT SLO on a cluster of 4 LLM servers, which translates into 64 and 96 RPS for clusters of size 8 and 12 servers respectively as confirmed by the experiment.

\subsection{Sensitivity to Rank Skews}
\begin{figure}[h]
\centering
    \centering
\includegraphics[width=0.99\linewidth]{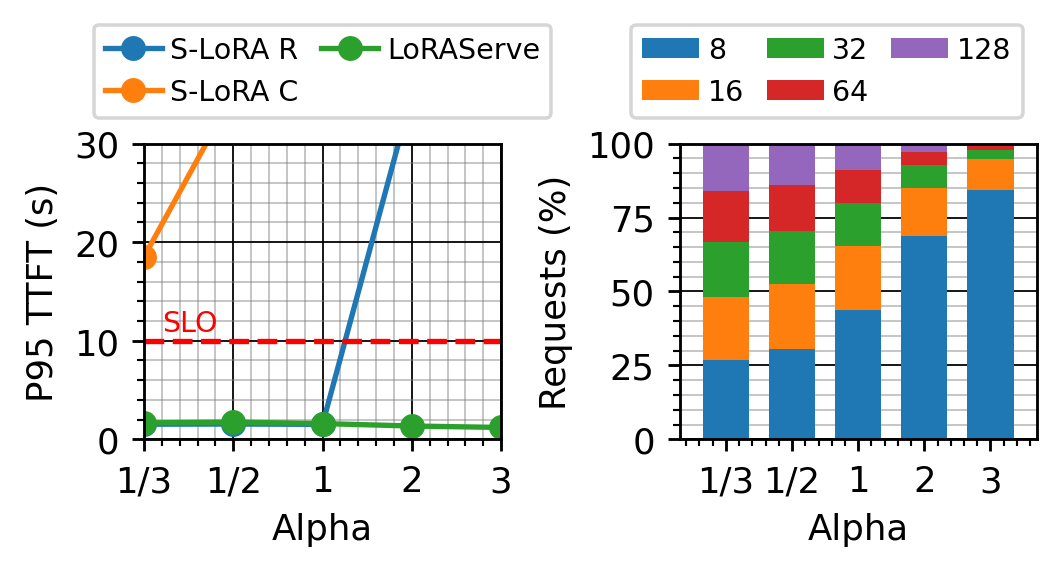} 
\caption{\textit{\textbf{Varying $\alpha$ in power law distribution for adapter popularity} on a 36 RPS Poisson arrival trace with 100 adapters (20 of each rank).}}
\label{fig:rank_skew}
\end{figure}

Fig \ref{fig:rank_skew} shows performance trends on traces with varying rank skews where the adapter popularity follows a power-law distribution ($y = x^{-\alpha}$) with smaller ranks receiving more requests \cite{chameleon}. We observe that although \sysname always shows tail TTFT performance well within SLO, upon increasing $\alpha$ from 1/3 to 3, the tail behavior slightly improves. This is because at lower $\alpha$, larger ranks receive a sizable share ($\geq 16\%$) of total requests whereas as $\alpha$ increases, the share presence of the largest rank in the trace diminishes and the trace is mostly dominated by small ranks as illustrated in Fig \ref{fig:rank_skew}. Thus, the performance gains can be attributed to smaller rank requests being naturally lighter. However, varying rank skews do not have any adverse effect on \sysnamex's performance unlike the baselines. 

The baselines, on the other hand, suffer with increasing skew in adapter popularity. Contiguous adapter placement shows poor tail performance and times out at higher skews due to load imbalance across nodes in the cluster as it only considers adapter ranks. Random adapter placement depicts performance similar to \sysname at smaller skews but times out as the skew increases. This is because, as a small number of adapters become highly popular, it is easy for a bad random allocation to happen, causing severe load imbalance.




\subsection{Sensitivity to Model Sizes}
\begin{figure}[h]
\vspace{-15pt}
\centering
    \centering
   \includegraphics[width=0.99\linewidth]{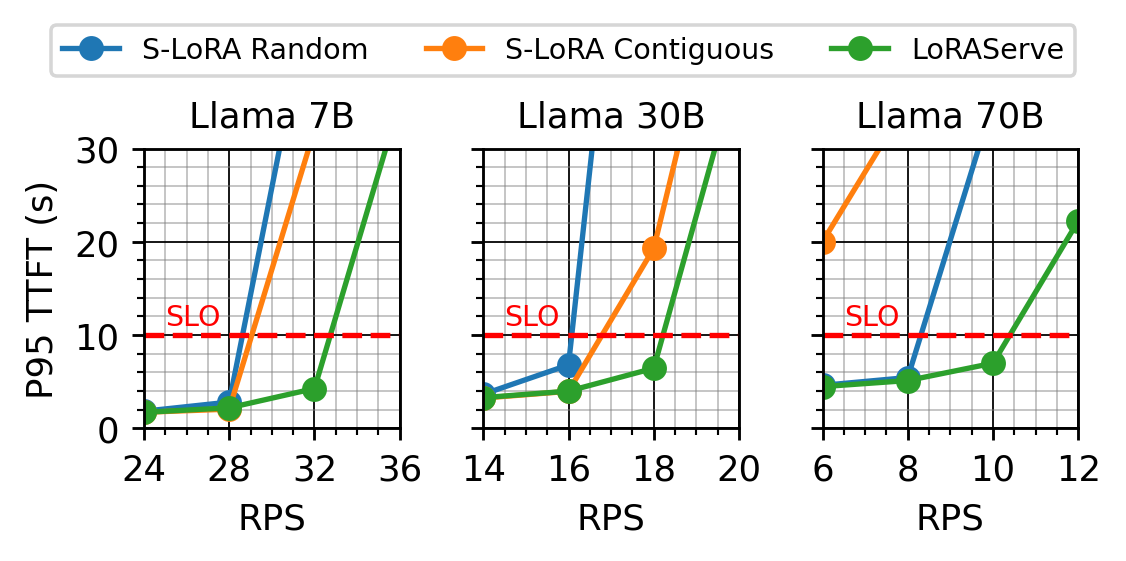} 
\caption{\textit{\textbf{Performance of \sysname using models of different sizes.}}}
\label{fig:model_sizes}
\vspace{-6pt}
\end{figure}

General purpose LLMs used in production at \orgname are generally larger in size, often upto 200B parameters. Fig \ref{fig:model_size} shows the performance of \sysname and baselines on larger models from the Llama family. We observe that the performance trends are quite similar to what is discussed in Section \ref{eval:perf} w.r.t. Llama 7B, except that the larger models serve lower throughput in general which is orthogonal to the design of \sysnamex. The baseline performance trends also translate similarly.


\subsection{Sensitivity to TP}
\begin{figure}[h]
\centering
\vspace{-10pt}
    \centering
   \includegraphics[width=0.99\linewidth]{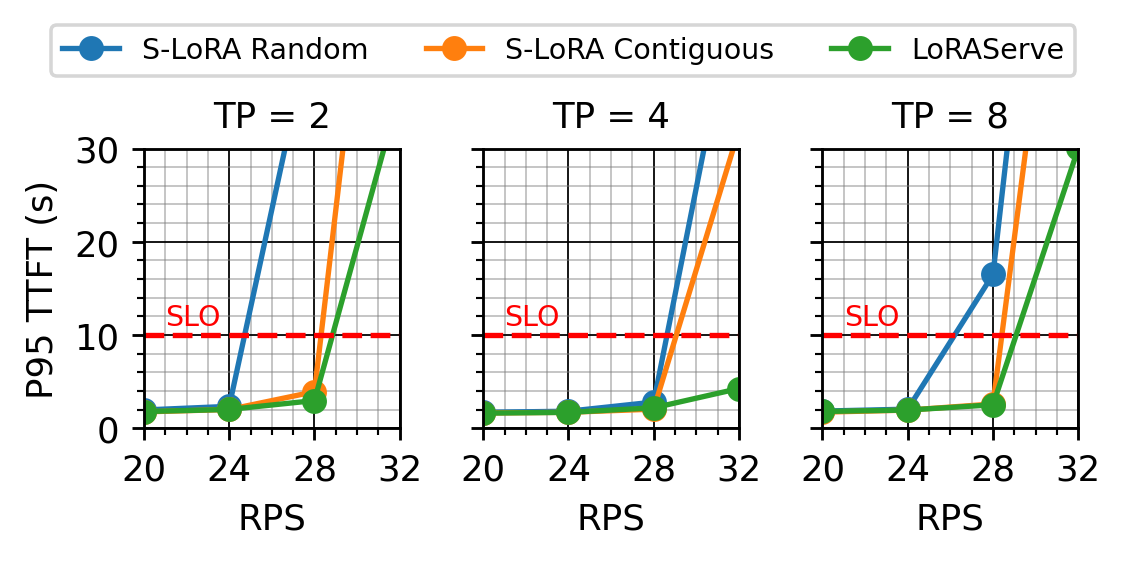} 
\caption{\textit{\textbf{Performance of \sysname using Llama 7B with different TP configurations.}}}
\label{fig:eval_tp}
\vspace{-8pt}
\end{figure}
Production LLMs are usually deployed under different configurations. We show that
\sysnamex's performance gains are orthogonal to the model's deployment configuration as it operates on a cluster scale. Fig \ref{fig:eval_tp} shows performance trends when models are deployed using various tensor parallelism configurations. In every case, \sysname complements the baselines, enabling higher throughput and lower TTFT.




\section{Related Work}
\subsection{LLM Inference}
Making LLM inference more efficient is the target of many recent studies. A vast corpora of recent papers have proposed hardware\cite{hardware-1, hardware-2}, software\cite{vllm}, kernel-level\cite{kernel-1, kernel-2} and system-level \cite{sageserve} optimizations for improving the performance\cite{orca,splitwise}, cost\cite{sageserve} and energy efficiency\cite{energy-1} of LLM Inference. These include optimizations in batching\cite{sarathi, batching-2}, scheduling\cite{orca, splitwise}, load-balancing\cite{loadbalancing, splitwise} and KV-caching\cite{kvcache-1, kvcache-2} among others.
While most of these works optimize for a monolithic LLM Instance deployed on a single GPU server, there are only a handful which consider cluster level optimizations. 
However, \sysname optimizes serving large scale serving of LoRA adapter workloads for a cluster of LLMs,
and can be used in conjunction with them to improve the cluster's utilization and performance.

\subsection{Parameter Efficient Fine-Tuning}
LoRA \cite{lora} and its derivatives such as QLoRA \cite{qlora} are the most popular of several parameter-efficient fine tuning \cite{peft} techniques \cite{prefix-tuning, prompt-tuning, ia3, adamix, compacter}. LoRA allows for zero overhead during inference by merging the adapter weights into the base model, as suggested in the original work \cite{lora}. 

While this approach is suitable when serving a single adapter, it does not scale to multiple adapters, especially of varying ranks and traffic patterns where multiplexing of adapters becomes necessary. 
Several new inference techniques address this issue. Punica \cite{punica} and S-LoRA \cite{slora} develop special CUDA kernels to batch requests from different adapters and decouple base model and adapter computation. S-LoRA also unifies paging across KV cache and LoRA adapters stored on the GPU, extending the paging seen in serving systems like vLLM \cite{vllm}. dLoRA \cite{dlora} dynamically merges and unmerges adapters with the base model and migrates requests and adapters between replicas. Chameleon \cite{chameleon} presents adapter caching on the GPU and adapter-aware scheduling. Toppings \cite{toppings} uses the CPU for LoRA computation while adapters are being loaded to the GPU to improve TTFT.

\section{Conclusion}
Rank heterogeneity and memory pressure in multi-tenant LoRA serving lead to severe performance degradation and inefficient resource utilization. 
Guided by insights from production like skewed adapter popularity and rank-induced interference, \sysname dynamically allocates and migrates adapters based on their rank and demand, optimizing performance and alleviating memory capacity pressure. Across diverse workloads and configurations, \sysname delivers up to 2$\times$ higher throughput, up to 9$\times$ lower TTFT, and uses up to 50\% fewer GPUs under SLO constraints, while reducing adapter storage memory footprint by up to 16$\times$, consistently outperforming state-of-the-art systems.



\bibliographystyle{IEEEtranS}
\bibliography{refs}

\end{document}